\documentstyle[osa,manuscript]{revtex}  % DON'T CHANGE
\begin{document}                % INITIALIZE - DONT CHANGE
\title{Doping dependence of the superconducting leading edge gap in Bi2212 }
\author{R. Gatt}
\address{Physics Department, University of Illinois at Chicago, Chicago IL60607}
%
%
%
%
% \author{}   % Use this and the next line only if there is a second
% \address{Another University, etc.}  % address. (Remove the left % marks)
%
\maketitle
\begin{abstract}                % DON'T CHANGE THIS LINE
The paper compares the K-dependence of the superconducting gap in different doping ranges. 
The fine behavior of the leading edge gap indicates that the pairing susceptibility is peaked 
at special regions on the Fermi surface. These hot regions are found to be centered away from 
nominal "hot spots". This 
behavior is attributed to a feedback effect on the pairing boson. Identification is made through
 comparison with neutron diffraction results.
\end{abstract}
\section{Text}               % Introduction goes below.

The mechanism of high temperature superconductivity is one of the most important current
problems of condensed matter physics, and a detailed description of the interaction which
leads to pairing is still lacking.
In common with most metals, the cuprate normal state
exhibits a Fermi surface \cite{phys report} . Phase sensitive experiments, 
\cite{Wollman,Tsuei} indicate that the superconducting order parameter exhibits a change of sign
around the Fermi surface \cite {problem}. Ding et. al. have shown that 
for nearly optimally doped Bi2212, the energy gap, determined from the peak position of the spectral 
function at the Fermi level, varies with
angle around the Fermi surface as $\Delta(\phi) = \Delta(0)Cos(2\phi)$ 
(first order d-wave),\cite{Ding} where $\phi$ is the 
Fermi surface angle measured from the Y-M direction of the Brillouin 
zone. In this report, the energy gap is determined from the measured 
Leading Edge Gap (LEG), for reasons to be detailed below.
The doping dependence of the LEG results indicates a pattern of 
behavior that provides insight into the superconducting pairing mechanism. \\
A general discussion on the electronic structure of the cuprates was given 
by Shen and Schrieffer\cite{S&S}. They have stressed the difference in the
 line shape and doping dependence
 of the spectral function along the $\Gamma - M$ direction (parallel to the 
 Cu-O bond) vs. along the $\Gamma - Y$ direction
  (parallel to the Cu-Cu direction). They suggested that such behavior could
   arise from electronic scattering 
  peaked at $Q=(\pi,\pi)$. They further suggested this scattering mechanism to 
  evolve into pairing susceptibility peaked 
  at $Q=(\pi,\pi)$ at low temperatures. This work provides an evidence 
  for a similar general behavior with some differences. \\
 
   Figure 1 is a comparison between the leading edge of a spectrum taken on 
    the Fermi surface of Bi2212
   well below $T_{C}$
   and the Fermi edge obtained from 
   a freshly evaporated Ag film. The zero line denotes the Fermi 
   level. The experimental details are published elsewhere\cite{Gatt}\\
   
   While the peak position method \cite{Hong} of determining the gap value 
   is considered rigorous at the Fermi 
   surface\cite{Anderson}
   and allows for inclusion 
   of energy and k-resolution, the LEG is chosen here on purpose as a 
   measure of the superconducting gap for the following reasons: 1. 
   The fitting procedure increases the error bars to a value of $\pm$ 
   3 meV\cite{Ding}, while the LEG method allows to push the 
   measurement to its current technological limit of $\pm$ 1 meV as 
   was shown by Ding et. al.
   \cite{Hong}. Figure 3 there demonstrates two points that are 
   important for our discussion: the ability to distinguish such 
   small shifts of the LEG and the rigidity of the LEG value to 
   the fit. This is easily understood by comparison of the sharp 
   leading edge and the shallow maximum (see Fig. 1).(The sharp feature at the 
   peak is a result of noise, and the peak position is determined by 
   fitting the spectrum to a given spectral function). 2. The LEG method 
   allows for direct comparison of the 
   superconducting gap and the pseudogap where the peak position 
   method is inapplicable. 3. The case of the high 
   $T_{C}$ cuprates is special since below $T_{C}$ appears a narrow energy region 
   close to the chemical potential where the imaginary part of the self energy is 
   substantially reduced \cite{Randeria} and the 
   spectral features approach resolution limit. The leading edge is 
   definitely in that region. The observed peak is at higher binding energy 
   where it is more likely to be affected by renormalization (due to 
   the finite energy resolution). 4. We are not interested in this study in the 
   absolute value of the superconducting gap (the LEG value is 
   numerically different than the nominal gap value), but in the relative value at different 
   points on the Fermi surface. (Thus in Fig.2 energy gap values for 
   different doping ranges are normalized to the maximum gap value). 
   In conclusion, It is not claimed here that any of the methods is 
   wrong, but that the larger error bars resulting from the peak fitting 
   procedure may screen finer details revealed by the shift of the 
   leading edge.\\

   Fig. 2 presents the angular
    dependence of the superconducting LEG in different doping regions. Zero degrees correspond to
    the $(\pi,\pi)-(0,\pi)$ direction of the Brillouin zone. The data 
    is presented in this unusual way to focus on the region of high 
    gap value. This region will be called the hot region\cite{MP}. Results are 
    reflected w.r.t. the $(\pi,\pi)-(0,\pi)$ direction (Y-M).
    The open diamonds are the results of Gatt et. al.\cite{Gatt}. 
    The filled diamonds are the
     LEG results of Ding et. al. on 
     87K slightly overdoped samples.\cite {Ding,Corsica}. The open circles 
     are LEG values for underdoped 75K samples\cite{Mesot}. There are 
     three regions to distinguish. The hot region close to Y-M 
     displays a flat behavior of the LEG. To avoid any connection to a 
     specific model we draw a straight line through the points in 
     this region. With increasing Fermi surface angle the LEG value
     drops sharply towards the node. Again, we draw a straight line 
     through the points in that region. The third region is the region 
     close to the node at 45 degrees. Here the gap drops less sharply 
     with decreasing slope as function of reduced doping\cite{Mesot}\\
     We try to use this data to identify the pairing boson. While 
     phonons and lattice anomalies correlate somewhat with the 
     transition temperature\cite{Egami,Keller}, spin fluctuations and 
     other electronic excitations measured by inelastic neutron 
     diffraction display a clear behavior that correlates 
     well with $T_{C}$\cite{Mason,Fong,Bourges,Mook}. We are looking for 
     a correlation between the gap spectrum and the spectrum of a 
     given boson. Spin fluctuations are characterized by a spin 
     susceptibility spectrum peaked at $Q=(\pi,\pi)$\cite{Bourges}. 
     The width of this spectrum is a function of doping with 
     (increasing width with increasing doping) in accordance with the 
     decrease of the magnetic correlation length.
     Several theoretical efforts were done to correlate this behavior 
     with superconductivity\cite{Moriya,MP,Chubukov}. Within this approach 
     it was suggested by Abanov et. al.\cite{ACF}that the angular dependence
     of the gap will reflect the q-dependence of the spin susceptibility with the magnetic 
    correlation length as a parameter. The set of data in the 
    overdoped range indeed follows that behavior\cite{Gatt}. The doping 
    dependence displays different behavior. At reduced doping the k dependence is 
  different due to the presence of  a transition to a 
    pseudogap state in addition to the superconducting 
    transition and a magnetic correlation length cannot be extracted 
    in the same way
    as in the overdoped case\cite{ACF}.
    To avoid any connection to a specific model, only straight lines 
    are used in the regions of flat behavior and rapid increase as 
    explained above. 
    We take the intersection of the straight lines as a measure of the 
    extension of the hot regions. We see that indeed the hot regions expand 
    with increasing doping in accordance with the decrease of
       the magnetic correlation length. Normalized to the extension 
       of the hot regions at optimal doping (OD87K) the extension in 
       the other doping regions is about 1.6 for the OD65K samples and 
       0.3 for the UD75K samples, in good agreement with the variation of the magnetic 
       correlation length with doping.
       Looking at the qualitative behavior, we realize two 
       important facts: 1. The extension of the hot regions decreases 
       continuously with decreased doping. 2. At all doping ranges the 
       hot regions are centered at $\phi = 0$. \\
       Fig. 3 sketches the Fermi surface\cite{Gatt} of overdoped 
       Bi2212.\cite{superlattice} The location of 
       hot spots $(K_{hs})$ is given by the intersection of the square with Q as 
       its side and the Fermi surface.\cite{energy} The intersection 
       of the straight lines in figure 2 is designated $K_{co}$, the 
       cutoff vector for the hot regions. The data of Fig. 2 indicate that 
       the hot regions are centered at the intersection of Y-M and the Fermi surface
        and NOT at hot spots. These points are designated $K_{pc}$ 
        (pairing ceters) since they appear as 
        pairing centers below $T_{C}$. The arrow on the Fermi 
        surface denotes the direction of movement of $K_{co}$ with 
        decreasing doping towards 
        the pairing center. The Fermi surface itself changes of course in 
        that region but slowly\cite{FS PRL}(fig. 3 there). This affects 
        the position of the hot spot so that $K_{hs}$ and $K_{co}$
        move in opposite directions with the latter moving more rapidly. The UD75K data is very 
        illuminating in that respect: $K_{co}$ has passed already the 
        hot spot and the hot region doesn't include the hot spot at all. This is impossible
        within the spin fluctuation mechanism unless there is a 
        feedback effect
        in the superconducting state\cite{Lee,Levin,thesis}: the scattering which is responsible 
        for pairing should be peaked now at two values along 
        the $(\pi,\pi)$ direction. This supplies a simple test which 
        doesn't require any fitting procedure (note that the data
         of Fig. 2 indicate unambiguously the center of symmetry): If 
         indeed the pairing bosons are spin fluctuations then the split 
         mentioned above into two different scattering vectors below 
         $T_{C}$, should 
         be evident in the spin susceptibility spectrum. The inset to 
         Fig. 3. are the neutron diffraction results of Mook et. al. on Bi2212. 
         \cite{Mook} It is clearly seen that two different peaks 
         appear below 
         $T_{C}$. The 
         values, (as measured from the figure), of 0.88Q and 1.14Q for the 
         spin susceptibility peaks should be compared with the Fermi 
         surface scattering vectors. Mook et al. report Tc of 84K. The Fermi 
         surface scattering vectors obtained from the measured Fermi 
         surface of 87K Bi2212 give 0.83Q  and 1.16Q 
         Correspondingly. It shows less 
         than 5 
         percents difference between the scattering vectors obtained from 
         the energy gap spectrum and the position of the peaks in the bosonic spectrum.
          \\
         An important question is how to interpret the commensurate 
         resonance observed at higher energy in Bi2212 as well as in 
         YBCO. Brinkmann and Lee\cite{Lee} have recently analyzed the energy 
         and q-dependence of spin fluctuations observed by inelastic 
         neutron diffraction. Their conclusion is that the resonance 
         is pushed into the gap region due to the presence of van 
         Hove singularity in the imaginary part of the bare spin susceptibility 
         induced by a nearly flat particle-hole excitation energy 
         spectrum for fermions with relative wave vector $Q=(\pi 
         ,\pi )$. They stress that setting t'=0 in their dispersion 
         relation causes the rersonance to be severly broadened. We 
         therefore consider the {\it commensurate} resonance as 
         representing the high energy part of the spin fluctuation spectrum 
         which may be related but not directly to the low energy pairing boson. \\

         Complementary information can be found in the effect of high energy 
         electron irradiation on superconductivity\cite{disordered}. 
         While the maximum gap hardly changes after irradiation, Tc is 
         reduced substantially and $K_{co}$ shifts toward the 
         pairing center as can be seen from the measured LEG before 
         and after irradiation (Fig. 4 there). \\
        As was mentioned above, the LEG method allows for direct 
        comparison of the gap below and above $T_{C}$ in underdoped 
        cuprates. Since the feedback effect is expected to occur below 
        $T_{C}$, It would be instructive to check weather 
        the recovery of the symmetric $(\pi,\pi)$ scattering above 
        $T_{C}$ is reflected in the k dependence of the gap. Such 
        unusual opportunity is supplied in the pseudogap state. 
        k-dependence measurements of the gap were published by Ding 
        et. al\cite{pseudogap}. Fig 3 there displays gap values in 
        the superconducting and pseudogap states. Unfortunately the 
        error bars are large. Still, from the data points it seems 
        that the two samples in the superconducting state display flat 
        behavior of the hot regions, centered at a pairing center, while the 10K sample, which is in 
        the pseudogap state shows a kink very close to a hot spot 
        location as can be verified from the published Fermi surface of 
        15K Bi2212\cite{FS PRL}. Detailed measurements at the 
        superconducting and pseudogap states are needed to clarify 
        that point. \\

         Unfortunately there are no published data on 
         the detailed angular dependence of the gap in other 
         materials, but a similar split below Tc \cite{Mook} 
         is observed in YBCO. The Fermi surface of YBCO was only 
         partially measured 
         \cite{YBCO} but it was found to be very close to that of Bi2212. 
         The pairing centers on Bi2212 are at high symmetry points. 
         Therefore they are probably at the same points in YBCO. This 
         implies that the locations of the incommensurate peaks along 
         the $(\pi,\pi)$ direction in YBCO 
         should be close to those in Bi2212 as indeed measured by Mook et. 
         al.\cite{Mook}. More ARPES measurements on YBCO are needed 
         to allow for a quantitative comparison.
        
         While a sharp resonance at $(\pi, \pi)$ is not observed in La214, 
         the incommensurate  peaks are observed with a very similar 
         displacement\cite{Mason}. The fact that the incommensurate peaks are 
         observed in all of these materials at very similar q-points, that their appearance 
         coincide with $T_{C}$ and that the Fermi surfaces are 
         similar, indicate that the spin fluctuations are major pairing 
         bosons in all of these materials.

\section{Aknowledgements}
It is a pleasure for the author to thank P. Chaudhari, A. Chubukov, 
C.C. Tsuei, A. Leggett, S. Rast, M. Weger, R. Joynt, M. Rzowski,F. 
Himpsel, J. Mesot, H. Ding  and J. C. Campuzano for exciting and useful 
discussions. The author is grateful for the 
technical support from the staff at the 
Synchrotron Radiation Center, Stoughton, Wisconsin. Financial support 
was provided by the U.S. NSF. The 
Synchrotron Radiation Center is supported by NSF. 
Correspondence should be addressed to Rafi@src.wisc.edu
%
%REFERENCES

%%%%%%%
 \begin{figure}  % Please send figures with disk, or separately if
% if it is an e-mail submission. (Good photo or India ink drawing.)
 \caption{Comparison of the spectrum from over-doped BI2212 and 
 freshly evaporated silver film. The Leading Edge Gap (LEG) is the 
 distance between the curves at the point of half maximum. The inset 
 shows the two edges in an expanded scale. It demonstrates that 
 differences smaller than 1 meV can be measured. This fine behavior 
 is the subject of the paper. \\}
\caption{Angular dependence of the LEG in different doping ranges of 
Bi2212. Open diamonds: over-doped 65K. Filled diamonds: over-doped 87K. 
Open circles: under-doped 75K. Straight lines are linear fits to data 
at the regions of flat and rapid change. The intersections of lines 
give the location of $K_{co}$ in the different doping ranges. The 
pairing center is the point of zero degrees on the horizontal line. \\ }
\caption{ The Brillouin zone and the Fermi surface of Bi2212. The 
vector Q is the magnetic scattering vector. $K_{co},K_{hs}$ and $K_{pc}$ 
are the locations of the cut-off vector, the hot spot vector and the 
pairing center vector. See the text. The transition from the square with 
corners at M to the rectangle with corners at $K_{pc}$ is the 
feedback effect described in the text. The inset are the results of 
Mook et. al. displaying the incommensurate neutron diffraction peaks 
observed at low temperature in Bi2212. \\ }
 \end{figure}

% \begin{table}
 %\caption{Please place your table caption here.}
 %\begin{tabular}{lrcd} % In second brace, l = left, r = right,
% c = centered and d = decimal justification.
 %One&Two&Three&Four\\  % Separate items with &. End line with \\
 %\tableline % Creates a horizontal line.
 %One&Two\tablenote{footnote.}&Three&Four\\ % Place \tablenote{}
% after item to be footnoted.
 %\end{tabular}
 %\end{table}


\begin{references}
\bibitem{problem}
Experimental results which do not show this behavior remain 
unresolved. \cite{Chaudhari,Sun}
\bibitem{superlattice}
For simplicity, the superlattice umklapp Fermi surfaces are omitted. 
This has no effect on the results of this paper which is based on 
measurements in the Y quadrant at 21-22 eV photon energy.
\end{references}
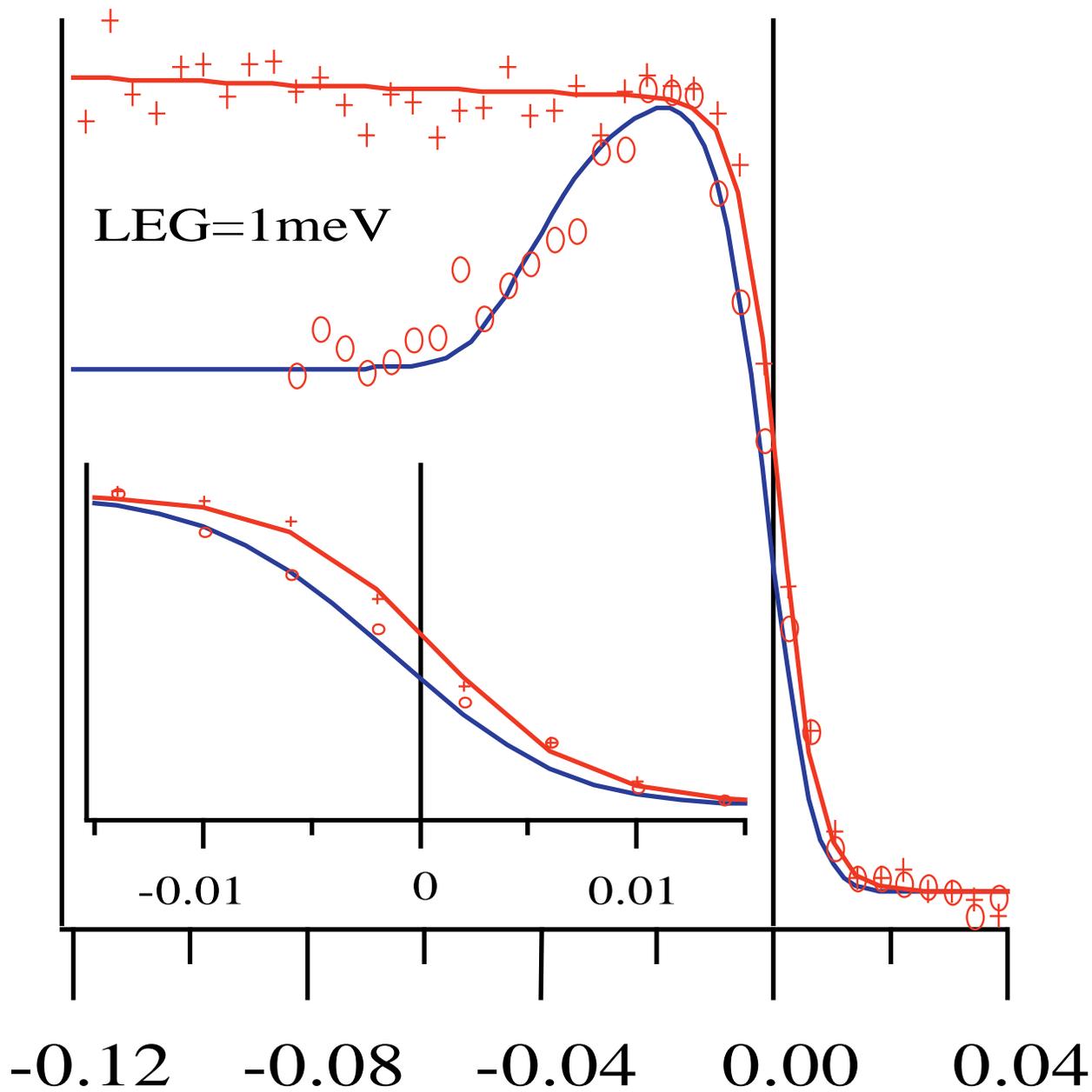
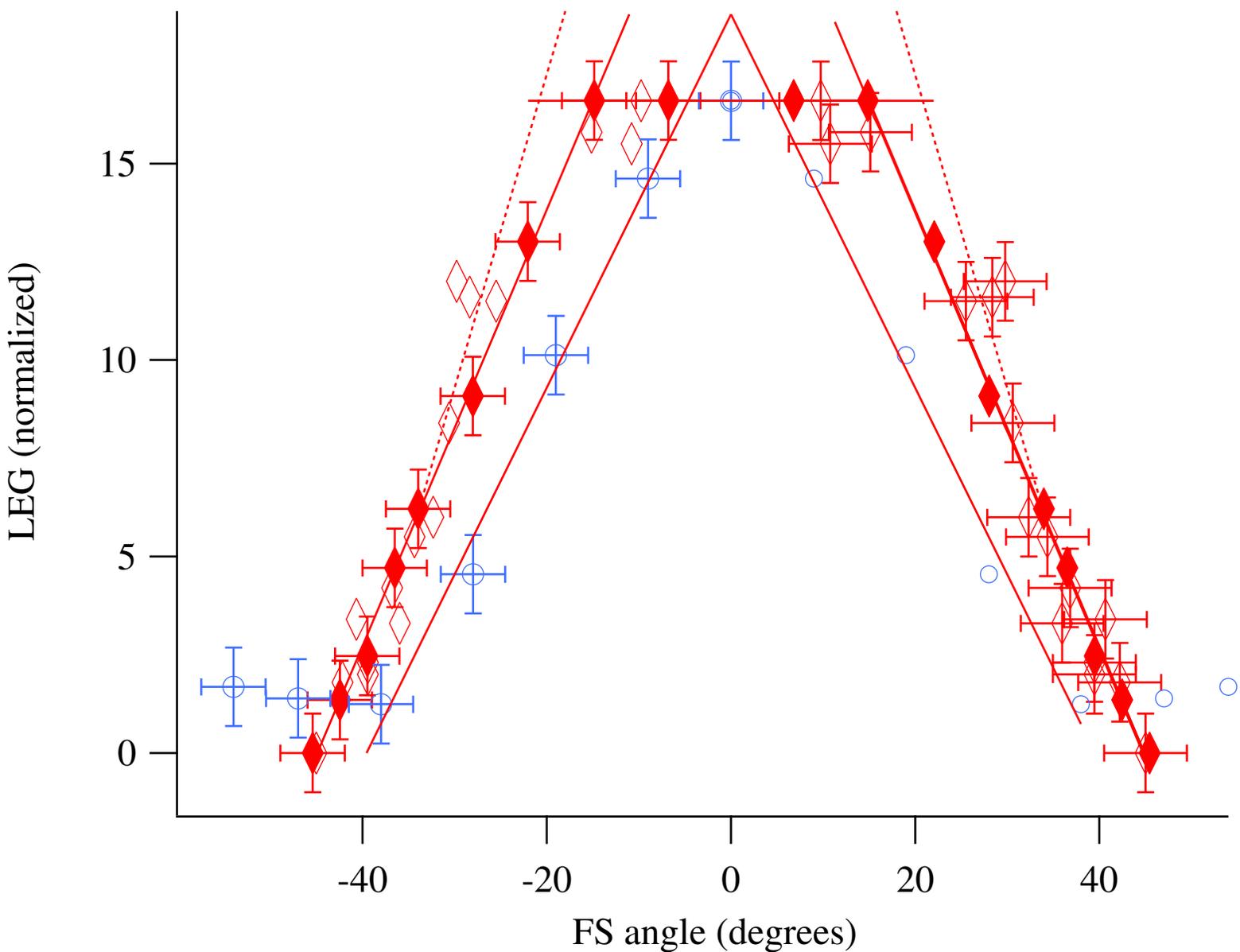
\end{document}